# Effects of surface vibrations on interlayer mass-transport: *ab initio* molecular dynamics investigation of Ti adatom descent pathways and rates from TiN/TiN(001) islands

D.G. Sangiovanni,[1,2*] A.B. Mei,[3,4] D. Edström,[2]
L. Hultman,[2] V. Chirita,[2] I. Petrov,[2,3] J.E. Greene[2,3]

[1]Atomistic Modelling and Simulation, ICAMS,
Ruhr-Universität Bochum, D-44801 Bochum, Germany

[2]Thin Film Physics Division, Department of Physics, Chemistry, and Biology (IFM),
Linköping University, SE-58183 Linköping, Sweden

[3]Departments of Materials Science, Physics, and the Frederick Seitz Materials Research Laboratory,
University of Illinois, Urbana, IL 61801, USA

[4]Department of Materials Science and Engineering,
Cornell University, Ithaca, New York 14853, USA

We carry out density-functional *ab initio* molecular dynamics (AIMD) simulations of Ti adatom (Ti$_{ad}$) migration on, and descent from, TiN <100>-faceted epitaxial islands on TiN(001) at temperatures $T$ ranging from 1200 to 2400 K. Adatom-descent energy-barriers determined via *ab initio* nudged-elastic-band calculations at 0 Kelvin suggest that Ti interlayer transport on TiN(001) occurs essentially exclusively via direct hopping onto a lower layer. However, AIMD simulations reveal comparable rates for Ti$_{ad}$ descent via direct-hopping vs. push-out/exchange with a Ti island edge atom for $T \geq 1500$ K. We demonstrate that the effect is due to surface vibrations, which yield considerably lower activation energies at finite temperatures by significantly modifying the adatom push/out-exchange reaction pathway.

*Corresponding author: Davide G. Sangiovanni, Tel. 0046 765506053, Fax 004613137568,
e-mail: davide.sangiovanni@rub.de, davide.sangiovanni@liu.se

# I. Introduction

Crystal growth is a complex phenomenon governed by an intricate interplay of thermodynamic and surface kinetic effects [1-3]. Control of nanostructural evolution during synthesis allows for tailoring material's properties to specific demands. In experiments, this is often achieved in a heuristic manner. However, the continuously rapid development of computational resources and efficiency combined with the reliability of first-principles techniques can provide atomic-scale knowledge of reaction pathways with corresponding rates at any temperature, pressure, or environmental condition of interest; an essential building block for advances in materials design, surface functionalization, and catalysis.

Half a century ago, direct experimental investigations by G. Ehrlich [4,5] and kinetic modeling by R.L. Schwoebel [6] revealed that a broken two-dimensional periodicity, caused by the presence of island step-edges, strongly modifies the local energy landscape probed by adatoms migrating on metal surfaces. Contrary to intuition, adatoms travelling toward descending lattice steps can be reflected by a potential-energy barrier, the Ehrlich-Schwoebel (ES) barrier, which prevents them from hopping onto lower layers. Subsequently, using field-ion microscopy [7,8], Ehrlich demonstrated that adatom incorporation at descending steps can also follow a pathway involving push-out and exchange with an island-step atom; that is, the adatom moves downward to replace an edge atom which is pushed to a position adjacent to the step [9].

More recently, Giesen and coworkers observed that the rates of Cu island decay on Cu(111) greatly increase when the separation distance $d$ between island and terrace step-edges is smaller than a critical value $w_c \approx 14$ Å [10,11]. The effect was attributed to ES barriers becoming vanishingly small for $d < w_c$, in accordance with theoretical models which showed that quantum confinement of electronic surface states in metals occurs for terrace widths $< w_c = \lambda_F/\sqrt{2}$, where $\lambda_F$ is the Fermi-electron wavelength [12]. However, after debating whether the ES barrier height is the only parameter which significantly affects adatom lifetimes on metal surface islands and terraces [13,14], Giesen et al. [15] and Morgenstern et al. [16] identified push/out-exchange descent as the mechanism which is primarily responsible for the onset of rapid interlayer transport on Ag(111).

Since Ehrlich's discovery in 1966, interlayer hetero- and homonuclear surface transport involving atomic exchange at descending lattice steps have been observed experimentally and demonstrated by *ab initio* calculations for material systems including W/Ir(111) [17], Pb/Cu(111) [18], and Pt/Pt(111) [19]. However, to our knowledge, interlayer transport via push-out/exchange has *not* been reported for compound surfaces, for which experimental determination of adatom and admolecule dynamics and quantitatively reliable evaluation of energy barriers for interlayer transport is further complicated by the presence of multiple chemical elements (see, for example, error bars exceeding experimentally-determined Ehrlich barrier values for TiN(001) in refs. [20,21]). Recent classical molecular dynamics (CMD) predictions indicate that push-out/exchange is the *sole* pathway for Ti adatom ($Ti_{ad}$) descent at <100>-oriented step-edges on TiN(001), a compound crystal surface [22]. In direct contrast, preliminary 0 K density-functional theory (DFT) calculations, yielding considerably smaller (by an order of magnitude) energy barriers for $Ti_{ad}$ direct-hopping onto the terrace, suggest that downward hopping is the dominant pathway for $Ti_{ad}$/TiN(001) interlayer transport over <100> steps.

In this work, we provide *ab initio* molecular dynamics (AIMD) verification of $Ti_{ad}$ push-out/exchange at <100>-oriented single-atom-high steps on TiN(001). NaCl-structure (*B*1) TiN is a model-system for transition-metal (TM) nitrides, a class of technologically important ceramic materials characterized by outstanding combinations of mechanical [23-27], physical [28,29], optical [30,31], electrical [32], and catalytic properties [33,34]. In addition, the properties can be tuned by varying TM compositions in alloys [32,35-38] and/or the N/TM ratio [39-43] over wide ranges, while preserving the cubic phase. AIMD simulations are used to separately determine the effects of varying the island area and temperature on adatom mobilities, as well as on $Ti_{ad}$ interlayer migration rates. We show that both the island size and surface vibrations contribute to modify the relative probability for direct-hopping vs. push/out-exchange reactions and that the collective atomic motion significantly alters the pathway of adatom push/out-exchange descent.

## II. Computational details

DFT calculations and DFT-based AIMD simulations [44] are carried out using VASP [45] implemented with the generalized gradient approximation (GGA) [46] and the projector augmented wave method [47]. All calculations employ Γ-point sampling of the Brillouin zone.

### A. AIMD simulations

In our AIMD simulations, *NVT* canonical sampling is performed by integrating the equations of motion at 1 fs time intervals, and controlling temperature via the Nosé-Hoover thermostat. At each time-step, the total energy is evaluated to an accuracy of $10^{-5}$ eV/atom using a plane-wave energy cutoff of 300 eV. Thermally-induced electronic excitations are accounted for by employing $k_BT$ electron-smearing. Visual molecular dynamics (VMD) software [48] is used to generate images and videos [49] of the results.

Intralayer and interlayer migration pathways with corresponding reaction rates are calculated for Ti adatom diffusion on, and descent from, single-atom-high square epitaxial TiN/TiN(001) islands with <100>-oriented edges and four Ti corner atoms. <100>-oriented edges are particularly relevant during TiN(001) film growth at relatively low N/Ti flux ratios and high (≥ 1200 K) temperatures, as observed by scanning tunneling microscopy and CMD simulations [50-52]. AIMD results are obtained at temperatures *T* between 1200 and 2400 K for two different island sizes.

$Ti_{ad}$ dynamics on larger 9×9-atom TiN epitaxial islands on a TiN(001) surface slab consisting of three layers (16×16 atoms per layer, ~12 nm² area, for a total of 850 atoms, Fig. 1a), are determined during eight runs (totaling ~110 ps) at *T* = 2400 K [53], well below the TiN melting point, $T_m \approx 3250$ K [54]. Smaller islands, comprised of 5×5 atoms, are placed on 12×12×3-atom TiN(001) slabs (Fig. 1b). Significantly less computationally-intensive than simulations of large substrates, AIMD modeling of the smaller 458-atom supercells allows following adatom trajectories at much lower temperatures, over calculation times which provide statistical accuracy. Thus, simulations of $Ti_{ad}$ dynamics on small islands are carried out at *T* = 1200, 1500, 2000, and 2400 K.

The total simulation time is ~150 ps (twice that required to obtain good statistics for reaction rates at 1200 K). A minimum of eight runs is performed at each $T$.

Prior to starting an AIMD simulation, the atoms in the TiN(001) island and substrate system are fully relaxed via DFT at 0 K. In all AIMD runs, atoms in the bottom slab layer are maintained fixed at relaxed positions. This does not affect the substrate temperature, which is determined by the translational degrees of freedom. Simulation boxes contain 15.3-Å-thick vacuum layers to minimize the interaction between vertical surface-slab replicas. The average in-plane Ti–N nearest-neighbor distance $d_{NN}$ in the terrace slab is obtained as a function of $T$ by rescaling the 0 K DFT+GGA value, $d_{NN}(0\ K) = 2.127$ Å [55], using the polynomial $d_{NN}(T) = a \cdot T^2 + b \cdot T + d_{NN}(0\ K)$. $a = 6.0868 \times 10^{-9}$ Å·K$^{-2}$ and $b = 9.3026 \times 10^{-6}$ Å·K$^{-1}$ are obtained by parabolic fitting of the experimental TiN lattice parameter variation due to thermal expansion as a function of $T$ [56]. Rescaling $d_{NN}$ is necessary to avoid spurious substrate strain effects on Ti$_{ad}$ migration kinetics [57]. Before initiating $NVT$ statistical analyses of adatom migration and island descent rates, thermal oscillations in the slab+adatom system are quickly stabilized with an isokinetic thermostat [58] for one ps. Ti$_{ad}$ initial positions are chosen stochastically to be one of the TiN island fourfold hollow (FFH) sites, the preferred Ti$_{ad}$/TiN(001) adsorption sites at both 0 K [59] and at elevated (> 1000 K) temperatures [60].

At each simulation temperature $T$, Ti$_{ad}$ migration rates $k(T)$ on TiN(001) islands are obtained for both <100> and <110> single jumps and compared to values estimated for diffusion on infinitely large TiN(001) terraces [60]. We note that 0 K DFT calculations by Tholander and coworkers [61] demonstrated that spin relaxation increases the local stability $\Delta E_{epitaxial}$ of Ti adatoms at metastable epitaxial TiN(001) surface sites from ~0.03 eV to ~0.10–0.15 eV. However, strong lattice vibrations at the temperatures employed in our present and previous AIMD simulations (≥ 1200 K) [60] effectively smear out the effects of Ti$_{ad}$ spin configurations on adatom surface transport. Local potential-energy minima are bypassed by itinerant adatoms since $k_B T \approx \Delta E_{epitaxial}$ [62]. Thus, the inclusion of Ti$_{ad}$ spin degrees of freedom in AIMD simulations would not significantly alter our calculated $Ti_{ad}/TiN(001)$ migration rates.

Interlayer descent rates for $Ti_{ad}$ direct hopping $k^{hop}(T)$ and push-out/exchange $k^{exc}(T)$ are obtained from AIMD runs as the total number of events observed divided by the total time spent by adatoms in FFH sites adjacent to island steps. Arrhenius linear regression is used to determine activation energies $E_a$ and attempt frequencies $A$ corresponding to both interlayer and intralayer adatom migration, evaluating uncertainties as detailed previously [63,64].

The temperature dependence of surface relaxation and island in-plane compression are determined by AIMD atomic coordinates averaged over (at least) 5 ps after thermalization. The effects of surface vibrations on the energy landscape probed by N and Ti atoms in 9×9 islands are evaluated by comparing average [001] vertical atomic coordinates $\bar{z}$ with the coordinate $z = \bar{u}$ corresponding to the minimum effective potential energy $U(z,T)$ at a temperature $T$. $U^{N,Ti}(z,T)$ is determined separately for N and Ti island atoms as described below. The vertical components $F_z^{N,Ti}$ of AIMD forces acting on N and Ti island atoms during ~7 ps are grouped in $z$ intervals $\Delta z$ = 0.04 Å and averaged within each interval: $F_z^{N,Ti}(z) = \langle F_z^{N,Ti} \rangle_{\Delta z}$. Thus, $U^{N,Ti}(z,T)$ is obtained as $U^{N,Ti}(z,T) = -\int_{z_{surf}}^{z} dz' \cdot F_z^{N,Ti}(z')$, for which $z_{surf}$ is the average [001] coordinate of atoms in the surface layer directly under the island. Phonon densities of states (PDOS) are obtained by Fourier-transforming AIMD velocity autocorrelation functions (VACF) of equilibrated supercells [65].

### B. Static DFT calculations

AIMD $E_a$ values are compared with $E_{a0}$ obtained via DFT at 0 K employing the nudged elastic band (NEB) method [66], with 400 eV as plane-wave energy cutoff. Initial and final stable states for NEB $Ti_{ad}$ interlayer diffusion are pre-relaxed via DFT conjugate-gradient energy minimization using AIMD configurations (prior to and after adatom descent) as input for the atomic coordinates. For $Ti_{ad}$ direct-jump NEB calculations, we also pre-optimize an AIMD transition-state configuration with the $Ti_{ad}$ located atop an island N edge atom, constraining $Ti_{ad}$ relaxation along [001]. Thus, 0 K migration energies are determined along minimum energy paths (MEP) sampled by at least seven, and up to a maximum of 13, NEB images.

Our statistical analyses are focused on determining the relative occurrence of $Ti_{ad}$ direct-hopping vs. push-out/exchange at island step-edges rather than at Ti island-corners, which are very rarely visited during AIMD simulations. This is consistent with our previous CMD observations (see Table 1 in Ref. [22]). We also note that island corners are blunted to minimize the step-edge formation energies under equilibrium conditions [67]. Thus, present DFT+NEB calculations of interlayer diffusion energy barriers $E_{a0}$ do not consider descents initiated in FFH sites at Ti island-corner positions.

DFT calculations are also used to quantify other important surface properties which influence TiN(001) surface dynamics and film growth, including island and surface relaxation, step-edge formation energies, and adatom adsorption and formation energies. Island step-edge formation energies $E_{\text{island step}}^{\langle 100 \rangle}$, the energy per unit length of <100> steps, are estimated as

$$E_{\text{island step}}^{\langle 100 \rangle} = \frac{E_{\text{slab+island}} - [n \cdot E_{\text{TiN bulk}}^{f.u.} + \mu(\text{Ti}_{\text{hcp}}) + E_{\text{surf}}^{(001)} \cdot A_{\text{slab}}^{(001)}]}{P_{\text{island}}^{\langle 100 \rangle}}. \qquad (1)$$

For the case of 9×9-atom islands, $E_{\text{slab+island}}$ (≈ –8091 eV) is the total energy of the relaxed supercell (849 atoms), $\mu(\text{Ti}_{\text{hcp}}) = -7.760$ eV/atom is the chemical potential of a Ti atom in a hcp Ti site, $P_{\text{island}}^{\langle 100 \rangle} = 65.086$ Å is the perimeter of the relaxed TiN island, $A_{\text{slab}}^{(001)} = 2317.443$ Å$^2$ is the total surface area of both upper and lower slab terminations, $E_{\text{surf}}^{(001)} = 0.078$ eV/Å$^2$ is the TiN(001) surface formation energy, $E_{\text{TiN bulk}}^{f.u.} = -19.526$ eV/f.u. is the energy per TiN formula unit in bulk TiN, and $n = 424$ is the total number of TiN formula units in our slab+island supercell. The addition of $\mu(\text{Ti}_{\text{hcp}})$ in Eq. (1) is required due to the fact that the islands in our supercell models contain one Ti atom more than N (see Fig. 1a). The effect of island corner relaxation on <100>-step energies is assessed by comparing $E_{\text{island step}}^{\langle 100 \rangle}$ values with $E_{\text{terrace step}}^{\langle 100 \rangle}$ results obtained for terrace <100>-step:

$$E_{\text{terrace step}}^{\langle 100 \rangle} = \frac{E_{\text{slab+terrace}} - [n \cdot E_{\text{TiN bulk}}^{f.u.} + E_{\text{surf}}^{(001)} \cdot A_{\text{slab}}^{(001)}]}{P_{\text{terrace}}^{\langle 100 \rangle}}, \qquad (2)$$

for which $E_{\text{slab+terrace}}$ is the total energy of a supercell comprised of 342 atoms (three TiN(001) monolayers + terrace, Fig. 1c) and $P_{\text{terrace}}^{\langle 100 \rangle}$ is the length of the two terrace steps in Fig. 1c. DFT step-energy results are converged to within an accuracy of ≈0.1 meV/atom with respect to k-point grid thicknesses.

Adatom formation energies $E_{\text{Ti}_{ad}}^{f}$ are calculated as

$$E_{\text{Ti}_{ad}}^{f} = E_{\text{slab+island+Ti}_{ad}} - E_{\text{slab+island}} - E_{\text{Ti}_{ad}}^{ads}/\text{terrace} - E_{\text{Ti}}, \qquad (3)$$

for which $E_{\text{slab+island+Ti}_{ad}}$ is the energy of the relaxed supercell with a Ti adatom adsorbed at an epitaxial position adjacent to the island edge, $E_{\text{slab+island}}$ is the energy of the relaxed TiN(001) slab+island system as given above, $E_{\text{Ti}_{ad}}^{ads}/\text{terrace}$ is $\text{Ti}_{ad}$ adsorption energy in a fourfold hollow site on an infinitely large TiN(001) terrace ($E_{\text{Ti}_{ad}}^{ads} \approx -3.3$ eV [50,60]), and $E_{\text{Ti}}$ is the energy of an isolated Ti atom (-2.275 eV [68]). Adatom adsorption energies on island $E_{\text{Ti}_{ad}}^{ads}/\text{island}$ are obtained using the expression

$$E_{\text{Ti}_{ad}}^{ads}/\text{island} = E_{\text{slab+island+Ti}_{ad}} - E_{\text{slab+island}} - E_{\text{Ti}}. \qquad (4)$$

### III. Results and discussion

### A. Structural, energetic, and vibrational properties of TiN islands and TiN(001) surfaces

Non-metal and metal atoms comprising the surface layer of *B*1 (001) TM nitrides and carbides relax vertically outward and inward, respectively. The effect is due to electron reorganization resulting from the broken lattice periodicity along the surface normal direction. The ripple amplitude $r_1$ quantifies the separation distance between the outermost (001) planes of non-metal and metal atoms. $r_1$ calculated for relaxed TiN(001) supercell slabs, 0.19 Å (Fig. 2b), is in agreement with previous DFT results (0.17 [59] and 0.18 Å [69,70]) and consistent with $r_1$ values experimentally determined for TiC (0.14 Å [71]), HfC (0.11 Å [72]), and TaC (0.20 Å [73]) (001) surfaces. Using the ideal $B$1 lattice sites as a reference, the (001) N plane protrudes outward by 0.12 Å, while the Ti plane relaxes toward the slab interior by 0.07 Å.

The ripple amplitude is not significantly affected by temperature. $r_1$ is slightly reduced from being 8.9% of the bulk nearest-neighbor spacing at 0 K (2.127 Å) to 8.4% at 2400 K. In contrast, the interlayer spacing $r_{1,2}$ (difference between the average vertical coordinates of N and Ti atoms in the surface layer and those in the second layer) progressively approaches the bulk Ti-N nearest-neighbor distance with increasing temperatures. While $r_{1,2}(0\ \text{K})$ is approximately 0.9% smaller than $d_{NN}(0\ \text{K})$, our parabolic fit of $d_{NN}$ and $r_{1,2}$ vs. $T$ data yields $r_{1,2}(T_m) \approx d_{NN}(T_m)$, Fig. 3. The variation in surface interlayer spacing increases faster with $T$ than that of the bulk Ti–N nearest-neighbor distance. At the melting point, $T_m \approx 3250$ K, we obtain increases of ≈5.3% for $r_{1,2}$ vs. ≈4.4% for $d_{NN}$ (inset in Fig. 3); both quantities reach average lengths of approximately 2.22 Å. The effect is due to anharmonic vibrations being stronger on the surface than in bulk [74]. As shown below, the rapid increase in surface vertical vibrational amplitudes significantly lowers the activation energy for adatom push/out-exchange at high temperatures.

DFT results show that relaxed square TiN islands contract in-plane and have slightly buckled <100> edges. N edge atoms adjacent to Ti edge-center atoms protrude outward by ~0.1 Å for both 5×5 and 9×9-atom islands (Fig. 2a). The [001] step-edge energy $E_{\text{island step}}^{\langle 100 \rangle}$ calculated for 9×9-atom islands (0.234 eV/Å) is in excellent agreement with previous experimental and DFT results (0.23±0.05 [20,75], 0.25±0.05 [76], and 0.238 eV/Å [77]). The island step energy, however, is 19% smaller than the value, $E_{\text{terrace step}}^{\langle 100 \rangle} = 0.289$ eV/Å, that we obtain for TiN(001) terrace steps. This indicates that the in-plane island relaxation has a significant effect on the energetics of <100> steps on TiN(001).

Along the diagonal of 9×9-atom islands, the average Ti–Ti distance is 2.876 Å; 4.4% smaller than the Ti–Ti interatomic spacing in bulk TiN (3.008 Å) calculated via DFT+GGA (Fig.

2a) at 0 Kelvin. The average nearest-neighbor distance along island medians (i.e., passing through the mid-points of opposite <100> edges), 2.053 Å, is 3.5% smaller than the 0 K Ti–N distance in the bulk (Fig. 2a). In distinct contrast to our results for the surface interplanar spacing $r_{1,2}$, for which Ti–N bond lengths along the surface normal direction become progressively closer to the bulk nearest-neighbor distance at temperatures approaching the melting point, lattice vibrations yield an island contraction which is more pronounced than at 0 K. At 2400 K, for example, Ti–Ti and Ti–N bond lengths along 9×9-atom island diagonals and medians are approximately 5.0% and 4.1% smaller than the corresponding bulk values.

The in-plane compression of 5×5-atom islands is even greater than that of 9×9-atom islands. At 0 K, island-diagonal Ti–Ti distances (2.804 Å) and island-median Ti–N distances (2.017 Å) are 6.8% and 5.2% smaller than the corresponding interatomic spacing in the bulk. Ti–Ti and Ti–N bond lengths determined by AIMD at 2400 K for smaller islands differ from bulk values (at the same temperature) by −7.7% and −6.3%. The relatively large difference in in-plane contraction of 5×5 vs. 9×9-atom islands indicates stronger bonding in smaller islands. This is consistent with the fact that average interatomic bond energies are greater for decreasing cluster sizes [78] (while average cohesive energies are smaller due to an increasingly larger fraction of unsaturated bonds [79]). These results are used below to discuss the difference in relative occurrence of adatom push/out-exchange vs. direct hopping from 5×5 and 9×9-atom islands.

Other important surface properties, which control both island coarsening and island decay kinetics [20,75,76,80,81] are the attachment/detachment and adatom formation energies. The energy required for an adatom to detach from an island edge onto the terrace can be partitioned into the adatom formation energy, surface migration energy, and the attachment/detachment energy (Fig. 4a). The magnitude of attachment/detachment energies and adatom formation energies depend on the nature of the interactions between an adspecies and a surface step. Ehrlich qualitatively assessed these properties by analyzing the differences in density of free admolecules away from, and proximate to, islands [82].

DFT evaluations of Ti$_{ad}$ formation energies $E^f_{Ti_{ad}}$ in terrace epitaxial positions adjacent to near-corner (2.06 eV) and near-center (2.20 eV) N edge atoms of larger islands, as well as $E^f_{Ti_{ad}}$ (2.15 eV) for adatoms adjacent to step edges of smaller islands (see Fig. 1a for definition of edge atom positions), are consistent with scanning tunneling microscopy results of Kodambaka and coworkers ($E^f_{Ti_{ad}} \approx 2.2 - 2.5$ eV [75]), obtained as the difference between the activation energies of TiN island decay (3.6±0.3 eV) and surface migration energies of adatoms or small admolecules (ranging from 1.1 to 1.4 eV depending on the nitrogen partial pressure [83]). It should be noted, however, that the experiments did not distinguish the type of adspecies attaching or detaching at TiN-island edges.

Calculated Ti$_{ad}$ adsorption energies on upper island edges decrease with decreasing island size. The adsorption energies of a Ti adatom in fourfold hollow sites along a descending step of 9×9-atom islands ($E^{ads}_{Ti_{ad}}$/island = −2.86 eV in FFH$_{near-corner}$, −3.05 eV in FFH$_{edge}$, and −3.11 eV in FFH$_{near-center}$ edge positions) are larger than the value obtained on 5×5-atom islands ($E^{ads}_{Ti_{ad}}$/island = −2.81 eV in FFH$_{edge}$). Overall, however, Ti$_{ad}$ adsorption energies in FFH sites on upper island edges are all smaller than the values obtained on infinitely large TiN(001) terraces ($E^{ads}_{Ti_{ad}}$/terrace = −3.2 eV [60], −3.3 eV [50]).

**B. Ti adatom diffusion on TiN/TiN(001) islands**

AIMD simulation results for Ti adatom dynamics at 2400 K on the larger TiN/TiN(001) islands yield <100> and <110> Ti$_{ad}$ jump rates (obtained for Ti$_{ad}$ migration over a total of ~82 ps) of $k^{<100>} = 3.80(\times 1.4^{\pm 1}) \times 10^{11}$ s$^{-1}$ and $k^{<110>} = 3.55(\times 2.3^{\pm 1}) \times 10^{11}$ s$^{-1}$. These are within, or close to, the numerical uncertainty of Ti$_{ad}$ migration rates on infinitely-large TiN(001) terraces [$k^{<100>} = 3.89(\times 1.8^{\pm 1}) \times 10^{11}$ s$^{-1}$ and $k^{<110>} = 1.79(\times 2.0^{\pm 1}) \times 10^{11}$ s$^{-1}$] extrapolated to $T = 2400$ K using $E_a$ and $A$ AIMD diffusion parameters from Ref. [60]. In contrast, Ti$_{ad}$ mobilites on smaller islands are larger than those previously determined for flat TiN(001) terraces at all temperatures. For example, at 2400 K, $Ti_{ad}$ jump rates on 5×5-atom islands are approximately twice that on terraces.

An Arrhenius linear regression of temperature-dependent $Ti_{ad}$ intralayer migration rates $k$ on 5×5 atom islands yields the following results: $k^{<100>}(T) = [8.1(\times 2.6^{\pm 1})$ THz$] \times \exp[-(0.48\mp0.07)$ eV$/(k_BT)]$ and $k^{<110>}(T) = [1.6(\times 2.4^{\pm 1})$ THz$] \times \exp[-(0.36\mp0.07)$ eV$/(k_BT)]$. In contrast, AIMD simulations of $Ti_{ad}$ migration on infinitely large terraces [60] yields $k^{<100>}(T) = [5.3(\times 1.7^{\pm 1})$ THz$] \times \exp[-(0.54\mp0.06)$ eV$/(k_BT)]$ and $k^{<110>}(T) = [1.5(\times 2.1^{\pm 1})$ THz$] \times \exp[-(0.44\mp0.07)$ eV$/(k_BT)]$. The temperature dependence of total adatom jump rates (i.e., including both <110> and <100> surface jumps) on smaller islands yield activation energies of 0.44±0.06 eV and attempt frequencies of $8.9(\times 1.6^{\pm 1})$ THz, vs. $E_a = 0.51\pm0.03$ eV and $A = 7.9(\times 1.2^{\pm 1})$ THz for TiN(001) terraces [60] (see Fig. 5). The enhanced mobility of Ti adatoms on smaller TiN(001) islands compared to terraces is due to in-plane island relaxation. AIMD activation energies and attempt frequencies determined for $Ti_{ad}$ intralayer transport on TiN(001) terraces and islands are summarized in Table I.

$Ti_{ad}$ FFH site occupation probabilities at 2400 K reveal that FFH corner positions are rarely visited: never observed on 5×5-atom islands and occupation probabilities are only ~5% on larger 9×9-atom islands. Edge FFH sites with the highest $Ti_{ad}$ sampling probabilities on 9×9-atom islands are: edge (~42%), near-center (~37%), and near-corner positions (~16%) (see inset in Fig. 1a for site definitions). In our AIMD simulations, itinerant Ti adatoms reaching corner sites on larger islands (~5%), or initialized at these positions for smaller islands [84], do not diffuse along step-edges nor directly hop onto the terrace. Instead, they quickly descend via push-out/exchange with a Ti corner atom.

## C. Ti adatom descent from TiN/TiN(001) islands

Homo- and hetero-nuclear interlayer adatom transport at surface steps has been extensively investigated both experimentally and theoretically for metallic systems including W/Ir(111) [17], Pb/Pb(111) [85], Pb/Cu(111) [18], Ag/Ag(110) [86], Pt/Pt(111) [19,87], and Au/Au(111) [88]. However, determination of descent pathways for ceramic materials such as TM nitrides is considerably more challenging due to the simultaneous presence of metal and non-metal elements which present a variety of bonding interactions (ionic, covalent, and metallic [55,89,90]) as well as surface morphology and N/TM stoichiometry which strongly depend on synthesis conditions [51,52,91-93].

Adatom incorporation at descending steps on solid surfaces occurs primarily via two competing reactions: adatom direct hopping onto a lower layer (Fig. 4b and 4c) and adatom push-out/exchange with a step-edge atom (Fig. 4d and 4e). In the case of homonuclear surface transport, the two reactions lead to the same atomic configuration. For $Ti_{ad}$/TiN(001), interlayer transport via push/out-exchange and direct-hopping produces distinguishable final states only by using different Ti isotopes. The magnitude of ES barriers directly affects the rate of adatom direct hopping and, indirectly, the rate of push-out/exchange descent. The ES barrier is defined as the difference between the migration energy on an island and the activation energy for direct-hopping onto a lower layer (Fig. 4a). ES barriers, recently imaged via friction-force microscopy on compound crystal surfaces [94], can be estimated experimentally from the difference in the rates of the growth (or decay) of two-dimensional adatom vs. vacancy islands [85].

All $Ti_{ad}$ direct hopping events recorded for large and small clusters are initiated with the adatom moving from a FFH site adjacent to the step-edge to a transition-state atop a N island-edge atom ($N_{edge}$) prior to descending into the nearest epitaxial position on the terrace. $Ti_{ad}$ descent via push-out/exchange requires concerted motion of the adatom with an edge Ti atom. The Ti adatom starts its downward migration from an island FFH edge site, thus causing an in-plane displacement of the nearest $Ti_{edge}$ atom. The reaction continues with the $Ti_{edge}$ adatom reaching an epitaxial site adjacent to the island edge and $Ti_{ad}$ occupying the vacated $Ti_{edge}$ position. The insets in Fig. 1 label $Ti_{ad}$ FFH island sites adjacent to a step-edge prior to adatom descent, as well as atop-N island-edge-sites (transition-states for $Ti_{ad}$ direct-hopping) and Ti island-edge atoms involved in push-out/exchange reactions. There are, for both $Ti_{ad}$ interlayer migration mechanisms, a total of four (two) unique descent pathways from 9×9- (5×5)-atom islands (Table II).

AIMD videos show that $Ti_{ad}$ migration across <100>-oriented descending steps on TiN(001)

primarily occurs via direct hopping. Nevertheless, our simulations reveal that push-out/exchange reactions become progressively more important with increasing temperatures and/or island sizes. As discussed below, we attribute these effects to interatomic bond strengths varying with island sizes as well as to island-atom vertical vibrational amplitudes becoming progressively larger with increasing temperature.

**C.1.** *0-Kelvin DFT+NEB results*. 0-K DFT+NEB results indicate that, irrespective of the island size and the upper edge position occupied by the adatom immediately prior to descent (Table II), $Ti_{ad}$ direct hopping over <100> step-edges has an energy saddle-point located midway between the FFH site on the island and the atop-N edge position. Figure 6 illustrates DFT+NEB pathways and energy landscapes for $Ti_{ad}$ direct-hopping onto the terrace from $N_{edge}$ sites of smaller islands. At 0 Kelvin, this reaction has an activation energy $E_{a0}^{hop}$ = 0.10 eV (see transition state labeled as position *b* in Fig. 6).

$E_{a0}^{hop}$ calculated for $Ti_{ad}$ hopping via atop-$N_{near-center}$ sites of 9×9-atom islands, 0.05 eV, is half that obtained for smaller islands, whereas direct hopping via $FFH_{near-corner}$ → atop-$N_{near-corner}$ trajectories has a much larger energy-barrier of 0.27 eV (see inset in Fig. 1a and Table II for interpretation of reaction pathways). Thus, applying the definition that the ES barrier is the difference between the *total* activation energy for adatom direct hopping and the adatom migration energy on the island (0.44 eV for $Ti_{ad}$ intralayer migration on smaller islands), ES barriers encountered by $Ti_{ad}$ at <100> step-edges would be *negative* for both 5×5- and 9×9-atom islands. Vanishingly small or negative ES barriers – which imply that adatoms travelling toward descending steps are more likely to be incorporated at terrace positions adjacent to step-edges rather than being reflected toward the island interior – and/or negative attachment energies (see Fig. 4a) – which favor condensation of itinerant adatoms at ascending terrace step-edges – promote surface smoothing during deposition via step-flow growth [95]. The existence of negative ES barriers for itinerant Ga adatoms has been proposed as a plausible explanation for stable homoepitaxial growth of GaAs crystals [95-97]. Our results confirm that the global activation energy for adatom descent is not necessarily larger than the activation energy for migration on an island as conventional schematic representations of Ehrlich barriers would suggest (see, for example, Fig. 4a of this work and figures in Refs. [20,21,75,98,99]).

For both larger and smaller islands, AIMD simulations reveal that $Ti_{ad}$ push/out-exchange is primarily initiated with the adatom in $FFH_{edge}$ sites (see Fig. 1a and 1b), the positions most frequently occupied by Ti adatoms at descending steps. Figure 7 shows that the 0-K NEB $E_{a0}^{exc}$ value for 9×9-atom islands, 0.91 eV, is ~18 times larger than the value obtained for $Ti_{ad}$ direct hopping, $E_{a0}^{hop}$ = 0.05 eV. The $E_{a0}^{exc}$ value obtained for 5×5-atom islands, 0.90 eV, is nine times greater than the 0-K activation energy calculated for adatom direct hopping. The large difference between $E_{a0}^{hop}$ and $E_{a0}^{exc}$ would suggest that push-out/exchange does not significantly contribute to interlayer adatom transport. However, to the contrary, AIMD results presented below demonstrate that the rate of adatom push-out/exchange is comparable with that obtained for direct hopping, even at moderately low temperatures (~40–50% $T_m$), that is, within the optimal temperature range for TiN(001) homoepitaxy [100,101].

**C.2.** *Finite-temperature AIMD results.* Theoretical investigations of diffusion processes and/or evaluations of reaction rates are often limited to 0 K estimates of minimum-energy paths and activation energies $E_a$ [61,102], whereas attempt frequencies are typically assumed to be of the order of a THz (lattice vibrational frequency) [103-105]. Moreover, $E_a$ values are also generally considered to remain constant with temperature. Such approximations, however, yield inaccurate predictions for the relative occurrence of interlayer-transport reactions at temperatures for which surface anharmonic vibrations become relevant. Lattice anharmonicity is also known to cause deviations from an Arrhenius temperature-dependent behavior for thermodynamic and kinetic properties in bulk systems. For example, it is known that vacancy formation [106,107] and migration (or diffusion, i.e. defect formation + defect migration) energies [108-110] may vary considerably from 0 K up to the melting temperature. Molecular dynamics simulations inherently resolve the problems mentioned above by integrating Newton's equations of motion for each atom at any temperature of interest. CMD/AIMD reaction rates for adspecies intra- and inter-layer migration [22,59,60,68,111] or desorption [63,112], as well as point-defects in bulk [110,113], can be employed in Kinetic Monte Carlo simulations [114] to efficiently probe the effects of precursors fluxes and ion-to-metal ratios on film growth modes.

An Arrhenius linear regression of AIMD $Ti_{ad}$ descent rates $k$ from 5×5-atom islands yield $E_a^{hop}$ = 0.10±0.06 eV and $A^{hop}$ = 1.1×1.9$^{±1}$ THz for direct hopping, and $E_a^{exc}$ = 0.33±0.12 eV and $A^{exc}$ = 1.0×2.1$^{±1}$ THz for push-out/exchange reactions (Fig. 8). Both 0-K DFT and finite-temperature AIMD activation energies and attempt frequencies determined for $Ti_{ad}$ interlayer transport across <100> island step-edges on TiN(001) are summarized in Table III. The fact that $E_a^{hop}$ evaluated at 1200 K ≤ T ≤ 2400 K matches (neglecting uncertainties) the DFT+NEB value calculated at 0 K suggests that lattice vibrations do not alter significantly the potential energy landscape probed by a Ti adatom during direct-hopping. In contrast, AIMD $E_a^{exc}$ values at 1500 K ≤ T ≤ 2400 K are approximately three times smaller than at 0 K. Consequently, while adatom direct-hopping rates extrapolated to high temperatures (using the 0 K DFT+NEB activation energy and an Arrhenius exponential prefactor of ~1.0 THz) would be reasonably close to AIMD values, the assumption that activation energies remain unaffected by lattice vibrations greatly *underestimates* finite-temperature $Ti_{ad}$ push-out/exchange rates; by nearly two orders of magnitude at 1500 K [$k_{0\,K \to 1500\,K}^{exc}$ = 9.5×10$^8$ s$^{-1}$ vs. $k^{exc}$(1500 K) = 7.8×10$^{10}$ s$^{-1}$ for smaller islands] and by more than one order of magnitude at 2400 K [$k_{0\,K \to 2400\,K}^{exc}$ = 0.1×10$^{11}$ s$^{-1}$ vs. $k^{exc}$(2400K) = 2.0×10$^{11}$ and 3.2×10$^{11}$ s$^{-1}$ for 5×5 and 9×9 islands]. Our observations are consistent with previous CMD results showing that surface vibrations are responsible for unexpectedly high Au adatom exchange rates in comparison to direct-hopping rates at Au(111) terrace steps [88] as well as for affecting or modifying diffusion pathways at finite temperatures, as in the case of Ag migration on Ag(110) [115].

As discussed above, DFT results show that the ES barrier encountered by a Ti adatom during downward diffusion across TiN <100> steps is negative, thus favoring surface smoothing and two-dimensional growth during synthesis at elevated temperatures. Negative ES barriers are consistent with total $Ti_{ad}$ interlayer transport rates being larger than $Ti_{ad}$ intralayer migration rates on terraces (Fig. 9).

Fig. 10 illustrates a typical AIMD $Ti_{ad}$ diffusion trajectory, during ~43 ps, on a 9×9-atom island at 2400 K. Starting in on FFH position in the island interior, the Ti adatom migrates along <110> and <100> directions to neighboring fourfold-hollow island sites. After approximately 10 ps of simulation time, the adatom lifts an island N atom onto the island, temporarily forming a TiN admolecule. The snapshots in the lower panels of Fig. 10 show the Ti adatom, laterally bonded to the former N island atom (yellow sphere), above the vacant anion site. After a few TiN admolecule in-plane rotations, the N atom returns to its original position in the island. At 42 ps, the Ti adatom reaches a $FFH_{near-corner}$ position, where it initiates a push-out/exchange with a $Ti_{edge}$ atom. The reaction, finalized within 1 ps, allows the adatom to descend to an $Ti_{edge}$ position, pushing the Ti island atom laterally toward an epitaxial site adjacent to the island edge.

Upward migration of N island atoms, which results in the formation of $TiN_x$ admolecules on an island, together with $x$ anion vacancies in the island, is induced by the synergistic effect of surface vertical vibrations and attractive $N/Ti_{ad}$ interactions. Temporary formation of $TiN_x$ admolecules on islands is relatively frequent at all $T$. For all N island atoms within a 2-Å-cutoff distance from an itinerant Ti adatom, the percentage of time τ spent above the island plane (i.e., at a vertical distance from the adatom smaller than 0.5 Å) increases dramatically with temperature. For smaller islands, τ = 1, 5, 17, and 20% for $T$ = 1200, 1500, 2000, and 2400 K, respectively. With larger islands τ = 18% for $T$ = 2400 K. The migrating Ti adatom can also permanently remove a N island atom from the island interior to form a stable TiN admolecule, thus leaving behind an anion vacancy. This occurs relatively often: 2 of 8 runs at 2000 K and 2 of 13 runs at 2400 K for 5×5-atom islands, and 2 of 8 runs at 2400 K for 9×9-atom islands. AIMD simulations reveal that TiN admolecule descent onto a terrace occurs either by directly hopping over the island step-edge or via a mixed mechanism in which one of the two atoms pushes an island atom (of the same element) out of the island edge while the other atom hops onto the lower layer. We have previously observed analogous TiN admolecule descent reactions at <100>-faceted TiN island steps during CMD simulations [111].

**C.3.** *<u>Effects of surface vibrations on adatom interlayer transport</u>.* The fact that AIMD push-out/exchange descent-rates from 5×5-atom islands are *not* vanishingly small (in relation to adatom direct-hopping rates), as would be expected based on the large difference between 0-Kelvin $E_{a0}^{hop}$ and $E_{a0}^{exc}$ values, is attributed to pronounced vertical vibrational amplitudes of island atoms, as clarified below.

In Sec. III.A, we showed that the TiN(001) ripple amplitude $r_1$ remains essentially constant with temperature. Thus, temperature-induced changes in surface/island undulation are expected *not* to produce substantial variations in Ti adatom dynamics or interlayer migration energies. In contrast, the surface interlayer spacing $r_{1,2}(T)$ increases with temperature faster than the TiN bulk nearest-neighbor distance $d_{NN}(T)$ (Fig. 3). AIMD simulations at 2400 K demonstrate that the vibrational amplitudes of surface atoms along the surface normal direction are up to ~70% larger than those within the (001) plane.

Phonon densities of states extracted by VACF allow quantifying the effects of temperature on TiN surface vibrations (Fig. 11). The PDOS results that we obtain via AIMD simulations of bulk TiN at 1200 K [116] are consistent with room-temperature phonon-spectra determined by neutron scattering measurements [117]. The density and dispersion of TiN(001) surface acoustic modes at 1200 K is essentially equivalent to that obtained for bulk at the same temperature (Fig. 11). TiN(001) surface vibrations, however, are characterized by softer optical modes, as demonstrated by the appearance of a PDOS peak at ≈12 THz and by the reduced PDOS at frequencies of 17 THz in comparison to bulk results. As expected, the reduced lattice periodicity has the effect of increasing the vibrational entropy and promoting anharmonicity. A comparison between the PDOS of TiN(001) at 1200 and 2400 K (Fig. 11) reveals that the elevated temperature causes a further softening in phonon modes with frequencies greater than 10 THz and produces an additional PDOS peak at low frequencies (≈4 THz).

The temperature-induced softening of surface phonons produces a shift in the average vertical coordinates $\bar{z}$ of N and Ti island atoms (solid blue lines in Fig. 12a and 12b) with respect to potential energy minima $\bar{u}$ along [001] (solid red lines). The potential energy landscape $U^{N}(z)$ probed by N island atoms exhibits a plateau (indicated by a black arrow in Fig. 12a) in correspondence of the positions occupied by the overlying Ti adatom. The effect, which originates from $Ti_{ad}/N_{island}$ attractive interactions, reflects the fact that N island atoms can be pulled onto the island to temporarily form TiN admolecules, as described in Sec. III.C.2. The $U^{Ti}(z)$ curve, instead, presents a smooth trend; the vertical vibrational amplitude of Ti island atoms (Fig. 12b) is much smaller than that observed for N atoms.

The PDOS and $U^{N}(z)$ results discussed above indicate that the large difference between finite-temperature ($E_a^{exc}$ = 0.33±0.12 eV) and 0-K ($E_{a0}^{exc}$ = 0.90 eV) activation energies arises from the changes induced by anharmonic surface vibrations on the chemical environment probed by Ti adatoms during push-out/exchange descent. The presence of a Ti adatom in a FFH site at an island edge enhances the out-of-plane vibrational amplitude of island N atoms close to the adatom. The effect, which becomes progressively stronger with increasing temperature, assists $Ti_{ad}$ push-out/exchange descent by weakening the bonding among $Ti_{edge}$ atoms and nearby N island atoms.

AIMD snapshots in Fig. 13 illustrate a typical AIMD reaction pathway for $Ti_{ad}$ push/out-exchange descent from a TiN(001) island. The descent from a $FFH_{near-corner}$ site on a 9×9-atom island is initiated with a Ti adatom lifting up the three closest underlying N island atoms and temporarily (< 1 ps) forming a $TiN_3$ admolecule on the island (Fig. 13a). The upward migration of N island atoms triggers the actual push-out/exchange by reducing the $Ti_{edge}$-atom bond coordination. $Ti_{edge}$ is displaced toward an epitaxial position adjacent to the island edge in concert with the downward translation of the adatom, while all N island atoms return to their former positions (Fig. 13b-d). 0-K DFT+NEB minimum-energy paths show, instead, that N island atoms are only slightly lifted from their positions during $Ti_{ad}$ push-out/exchange reactions (Fig. 7).

To summarize, the vertical vibration of island atoms promote push/out-exchange descent by significantly changing the adatom reaction path, in turn, resulting in smaller activation energies. Analogous to the effects of island vertical vibrations, we also considered the possibility that lateral vibrations of atoms at island step edges could contribute to facilitating the exchange process at high temperatures. However, for a given island size, the average in-plane Ti–N distance along island medians is not sensitive to temperature (see Sec. III.A). For example, with 9×9-atom islands, the equilibrium bond lengths between step-edge atoms and neighboring interior atoms remain 2.00±0.03 Å, irrespective of temperature. Therefore, for a given island size, we conclude that in-plane vibrations do not play a significant role in affecting $E_a^{exc}$ values at elevated temperatures.

**C.4.** *Effects of island size on adatom interlayer transport.* The result that push-out/exchange activation energies vary significantly with temperature due to pronounced surface vertical vibrational amplitudes is an important outcome of the present investigations. In addition, AIMD simulation results also reveal that the relative occurrence of Ti adatom push-out/exchange increases with the island size. While, indeed, the rate of $Ti_{ad}$ push-out/exchange from 5×5-atom islands is smaller than that obtained for direct-hopping at all investigated temperatures, $k^{exc}$ rates obtained for larger 9×9-atom islands are slightly higher than $k^{hop}$. We attribute this effect to an overall decrease in average bond strength in larger islands [78]. Stronger interatomic bonds within 5×5-atom islands, as reflected by a more pronounced in-plane contraction at all temperatures (see Sec. III.A), prevent bond breakage between Ti step-edge atoms and neighboring N island atoms, thus hindering push-out/exchange reactions.

Overall, AIMD results at 2400 K indicate that both adatom intralayer (Fig. 5) and interlayer (Fig. 8) mass transport on 5×5-atom islands occurs faster than on 9×9-atom islands.

## IV. Conclusions

AIMD simulations are used to probe Ti adatom intra- and interlayer migration on <100> faceted TiN/TiN(001) islands. Although 0-Kelvin *ab initio* activation energies determined for adatom direct-hopping are an order of magnitude smaller than those obtained for push/out-exchange reactions, AIMD results reveal that adatom push/out-exchange descent is a relevant pathway for interlayer mass transport on TiN(001) at moderate temperatures (approximately ~40–50% of the melting point). Analyses of finite-temperature vs. 0-K migration pathways, phonon densities of states, and effective potential energy landscapes along the surface orthogonal direction suggest that $Ti_{ad}$ push-out/exchange is promoted by surface anharmonic vibrations. The synergistic effect of large vertical island-atom vibrational amplitudes and attractive $Ti_{ad}$–N forces assist upward migration of N island atoms. This leads to the temporary formation of $Ti_{ad}N_x$ adspecies, which favors the displacement of Ti edge atoms due to reduced bond coordination with neighboring N island atoms. In addition, we show that adatom push-out/exchange descent is favored for larger island areas due to overall weaker interatomic bonds among island atoms.

## Acknowledgments

This research was carried out using resources provided by the Swedish National Infrastructure for Computing (SNIC), on the Gamma and Triolith Clusters located at the National Supercomputer Centre (NSC) in Linköping, and on the Beskow cluster located at the Center for High Performance Computing (PDC) in Stockholm, Sweden. Prof. Suneel Kodambaka is acknowledged for useful discussions. W. Olovsson and P. Münger at NSC, and H. Leskelä and J. Vincent at PDC are acknowledged for assistance with technical aspects. We gratefully acknowledge financial support from the Knut and Alice Wallenberg Foundation (Isotope Project No. 2011.0094), the Swedish Research Council (VR) Linköping Linnaeus Initiative LiLi-NFM (Grant No. 2008-6572) and Project Grants No. 2014-5790 and 2013-4018, the Swedish Government Strategic Research Area Grant in Materials Science on Advanced Functional Materials (Grant No.MatLiU 2009-00971 through Sweden's innovation agency VINNOVA), and the Olle Engkvist Foundation.

# Tables

| Ti$_{ad}$ intralayer transport at finite temperature (AIMD) | | | | | | | |
|---|---|---|---|---|---|---|---|
| 5×5 island | | | | Terrace | | | |
| <100> migration | | <110> migration | | <100> migration | | <110> migration | |
| $A$ (THz) | $E_a$ (eV) | $A$ (THz) | $E_a$ (eV) | $A$ (THz) | $E_a$ (eV) | $A$ (THz) | $E_a$ (eV) |
| 8.1×2.6$^{±1}$ | 0.48±0.07 | 1.6×2.4$^{±1}$ | 0.36±0.07 | 5.3×1.7$^{±1}$ | 0.54±0.06 | 1.5×2.1$^{±1}$ | 0.44±0.07 |

**Table I**. AIMD activation energies and attempt frequencies determined for Ti$_{ad}$ intralayer transport on TiN(001) terraces [60] and islands.

| Ti$_{ad}$ descent pathway | Direct hopping | | | | Push/out – exchange | | | |
|---|---|---|---|---|---|---|---|---|
| | 5×5 island | | 9×9 island | | 5×5 island | | 9×9 island | |
| | FFH site prior to descent | → N descent site | FFH site prior to descent | → N descent site | FFH site prior to descent | → Pushed Ti island atom | FFH site prior to descent | → Pushed Ti island atom |
| 1 | Corner | → Edge | Corner | → Near-corner | Corner | → Corner | Corner | → Corner |
| 2 | Edge | → Edge | Near-corner | → Near-corner | Edge | → Edge-center | Near-corner | → Edge |
| 3 | | | Edge | → Near-center | | | Edge | → Edge |
| 4 | | | Near-center | → Near-center | | | Near-center | → Edge-center |

**Table II**. Summary of symmetrically-unique Ti adatom descent pathways from smaller (5×5) and larger (9×9) TiN/TiN(001) square islands (see insets in Fig. 1 for definitions of atomic sites). All reactions end with a Ti atom (either an adatom or an island atom) in an epitaxial position adjacent to the island edge.

| Ti$_{ad}$ interlayer transport | | | | | | | |
|---|---|---|---|---|---|---|---|
| Finite-temperature (AIMD) | | | | 0 Kelvin (DFT+NEB) | | | |
| 5×5 island | | | | 5×5 island | | 9×9 island | |
| Push-out/exchange | | Direct hopping | | Push-out/exchange | Direct hopping | Push-out/exchange | Direct hopping |
| $A$ (THz) | $E_a$ (eV) | $A$ (THz) | $E_a$ (eV) | $E_a$ (eV) | $E_a$ (eV) | $E_a$ (eV) | $E_a$ (eV) |
| 1.0×2.1$^{±1}$ | 0.33±0.12 | 1.1×1.9$^{±1}$ | 0.10±0.06 | 0.90 | 0.10 | 0.91 | 0.05; 0.27 |

**Table III**. 0-K DFT+NEB, together with finite-temperature AIMD, activation energies and attempt frequencies determined for Ti$_{ad}$ interlayer transport across <100> step-edges of 5×5 and 9×9-atom TiN islands.

**Figures**

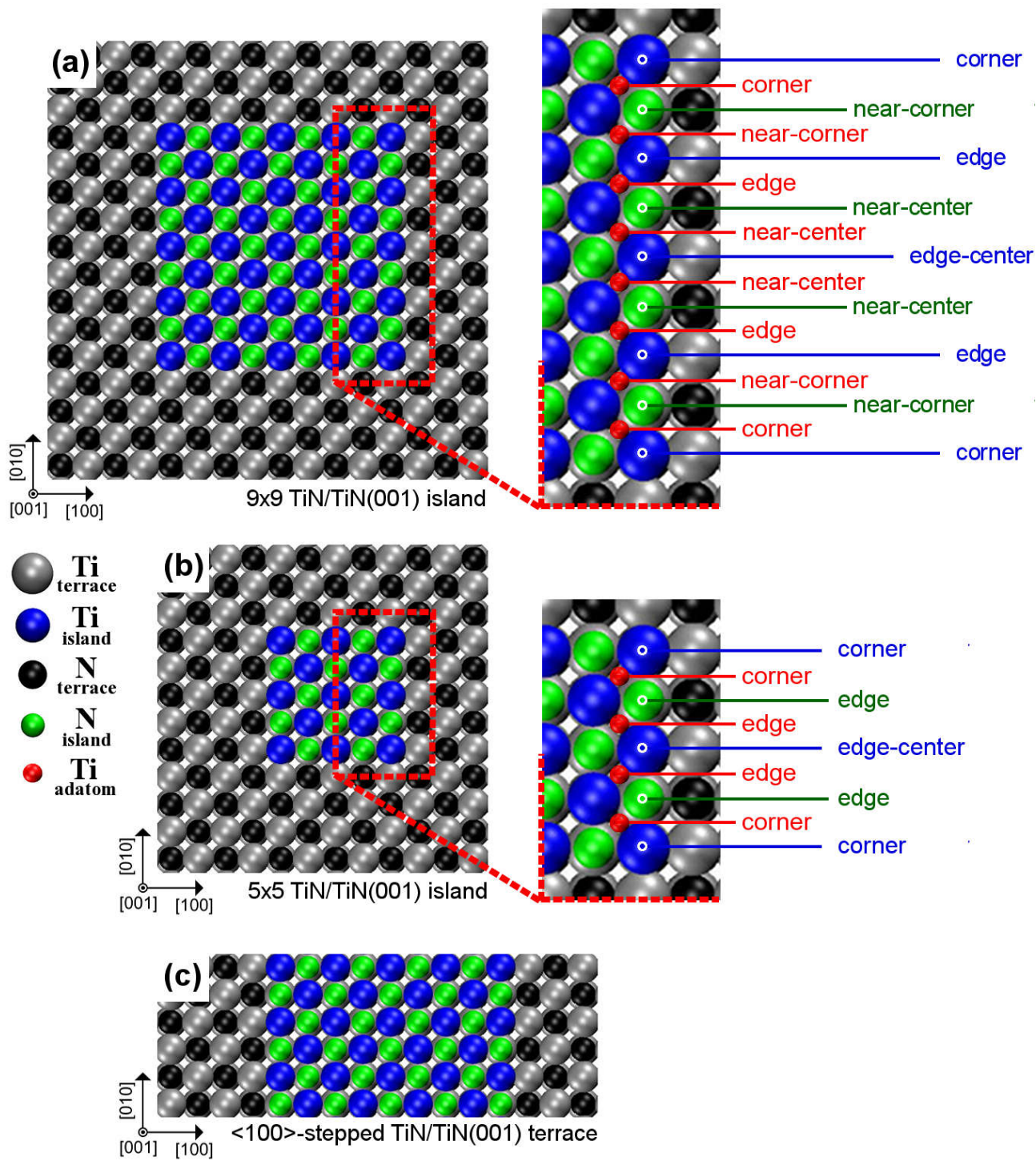

**Fig. 1**. Plan view of supercells used in AIMD simulations and DFT+NEB calculations. Note that the islands have fourfold symmetry: with mirror planes at <110> diagonals and <100> medians. (a) For large 9×9-atom islands there are: *(i)* four different FFH sites (red) that can be occupied by Ti adatoms at descending island steps; *(ii)* three nonequivalent Ti step-edge atoms (blue); *(iii)* three different N step-edge atoms (green). (b) For small 5×5-atom islands, there are: *(iv)* two different FFH sites (red); *(v)* two nonequivalent Ti step-edge atoms (blue); *(vi)* step-edge N atoms all have equivalent symmetry (green). (c) <100>-stepped TiN/TiN(001) terrace used for <100>-step energy DFT calculations.

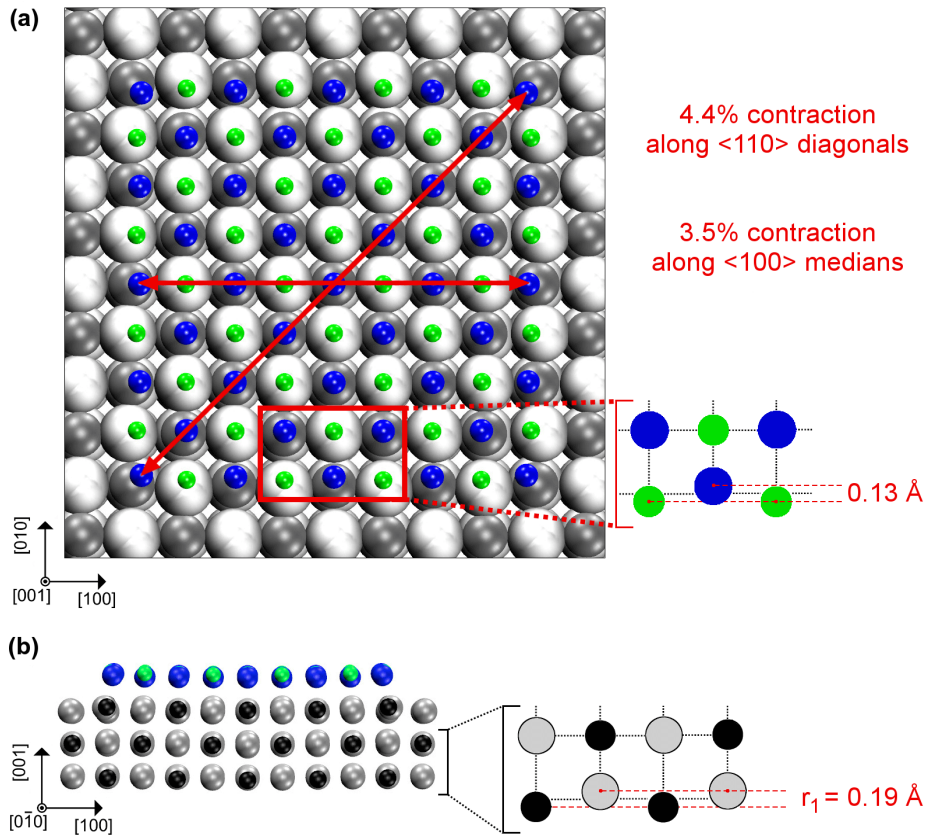

**Fig. 2**. DFT results for the relaxed substrate+island system. For better visualization of the in-plane island relaxation, terrace atoms are plotted in gray scale while island atoms (blue spheres = Ti, green spheres = N) are shown in reduced sizes.

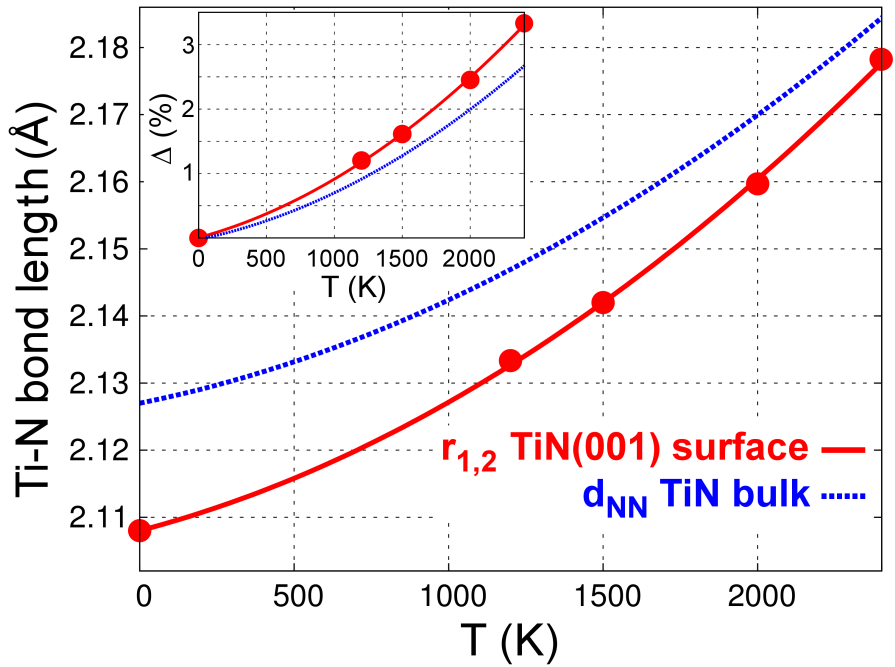

**Fig. 3**. TiN bulk and surface thermal expansion. The inset shows the normalized variation $\Delta = [\,(r_{1,2}(T) - r_{1,2}(0\,K))\,/\,r_{1,2}(0\,K)\,]$ in the $r_{1,2}$ distance (red filled circles) compared to the TiN bulk interatomic spacing (blue dotted line). $r_{1,2}$ data are fit (red line) with a polynomial $a' \cdot T^2 + b' \cdot T + r_{1,2}(0\,K)$, for which $a' = 7.0528 \times 10^{-9}$ Å·K$^{-2}$, $b' = 1.2098 \times 10^{-5}$ Å·K$^{-1}$, and $r_{1,2}(0\,K) = 2.108$ Å.

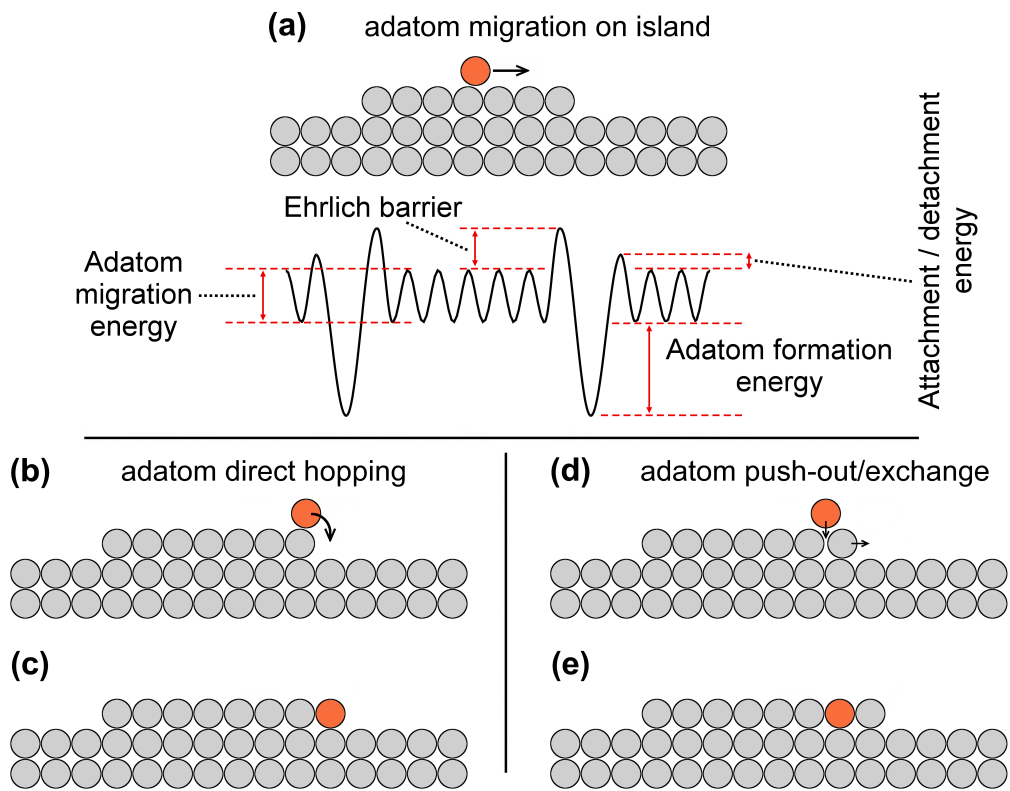

**Fig. 4.** Schematic illustration of (a) adatom migration on an island; (b) and (c) adatom direct hopping from island onto terrace; and (d) and (e) adatom descent via push-out/exchange with an island edge atom.

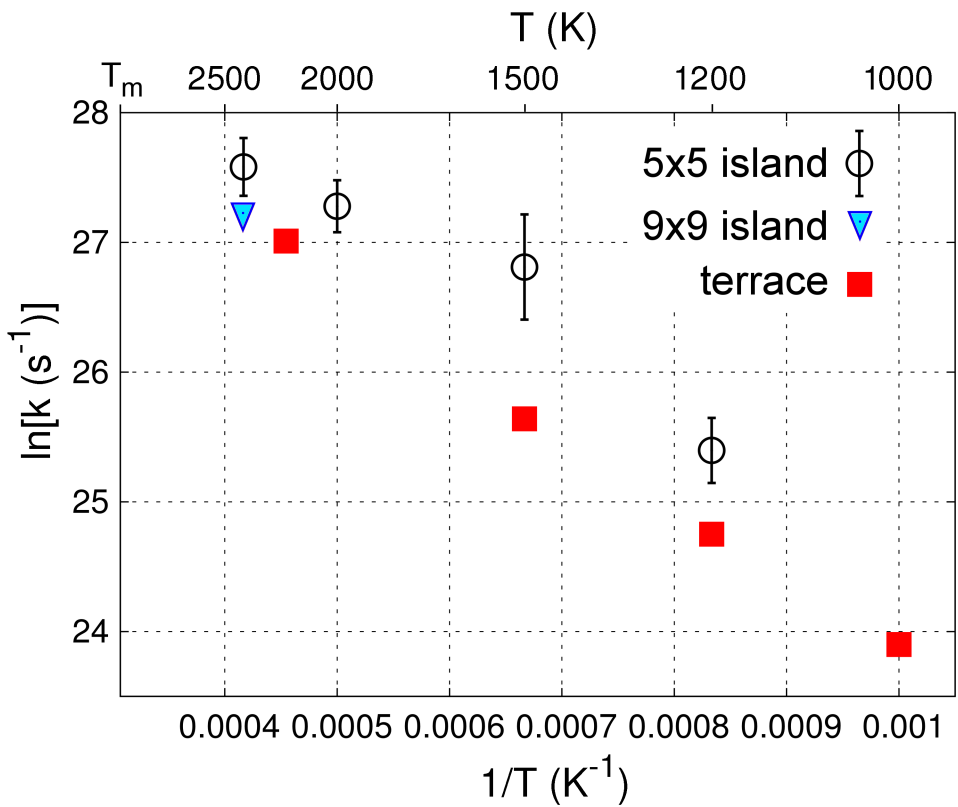

**Fig. 5.** Total Ti$_{ad}$ adatom jump rates $k$ on 5×5- and 9×9-atom TiN/TiN(001) islands vs. TiN(001) terraces [60]. The uncertainty bars for migration rates on larger island are smaller than the symbols.

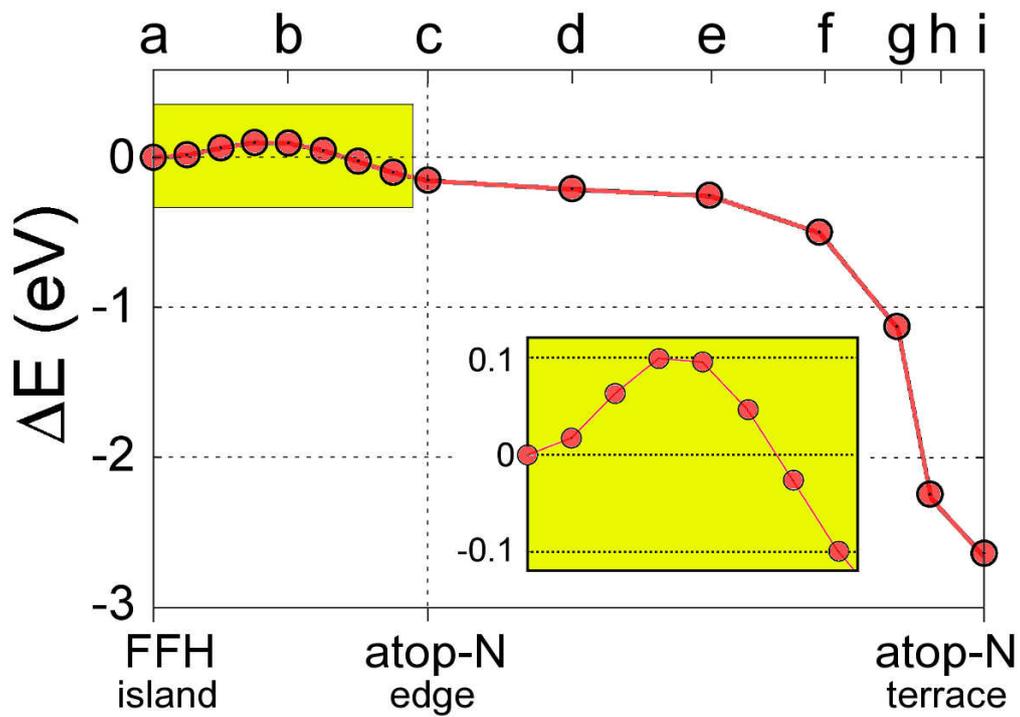
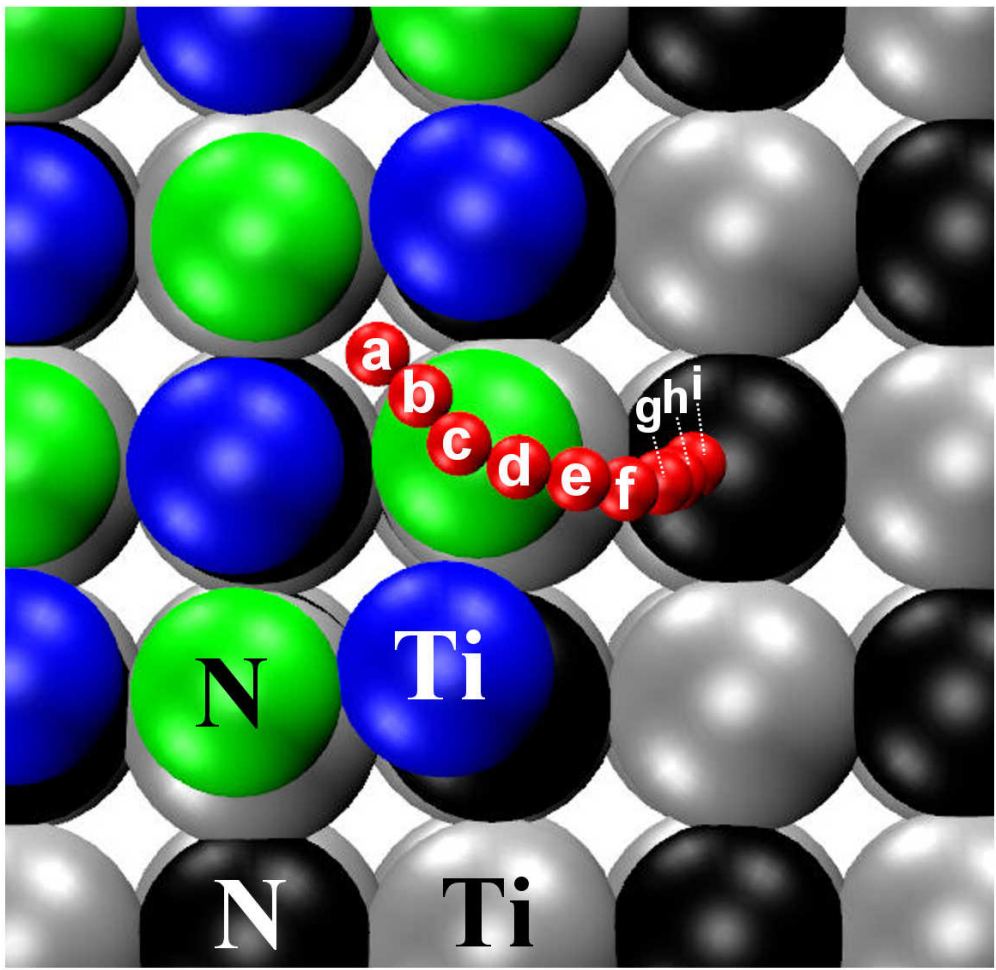

**Fig. 6**. Upper panel: DFT+NEB results for $Ti_{ad}$ direct hopping onto the terrace over N edge atoms on smaller (5×5 atoms) TiN/TiN(001) islands. Lower panel: Starting in a FFH island position (**a**), the Ti adatom descends by following a curved trajectory, first moving atop a N edge atom (**c**) then jumping onto the closest N TiN(001) terrace atom (**i**). The pathways for Ti adatom direct-hopping over <100> edges of larger TiN/TiN(001) clusters are equivalent to the one illustrated in this figure.

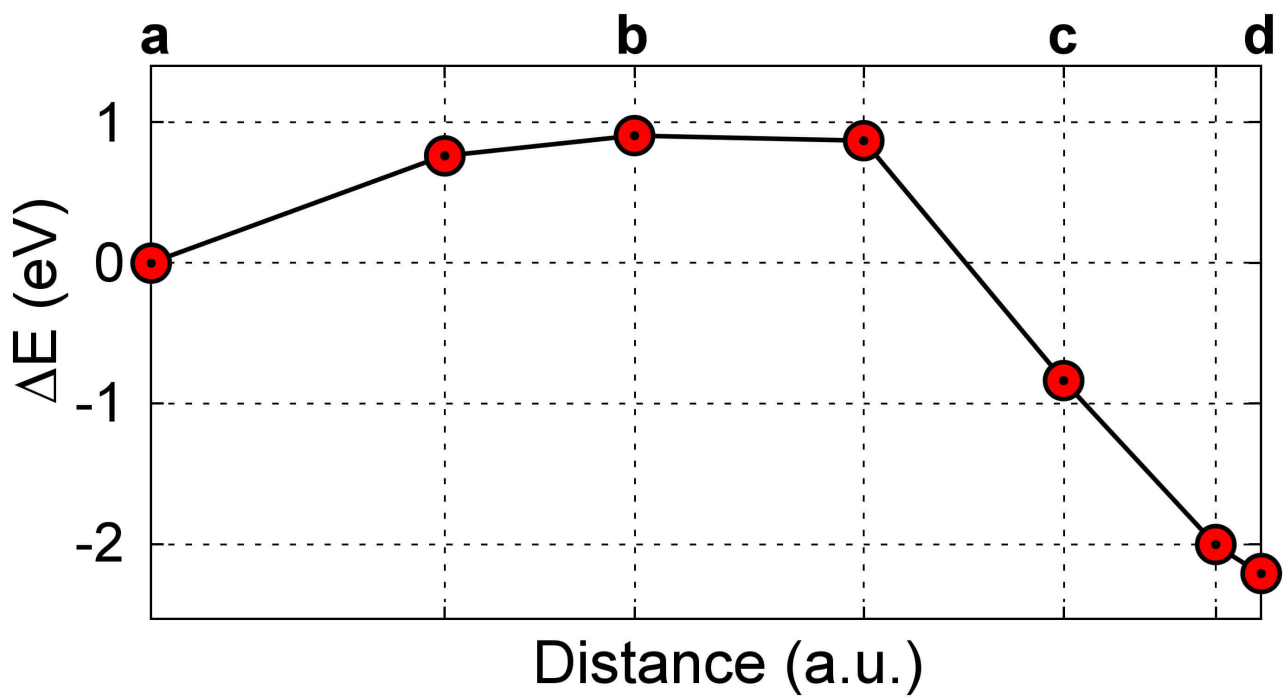

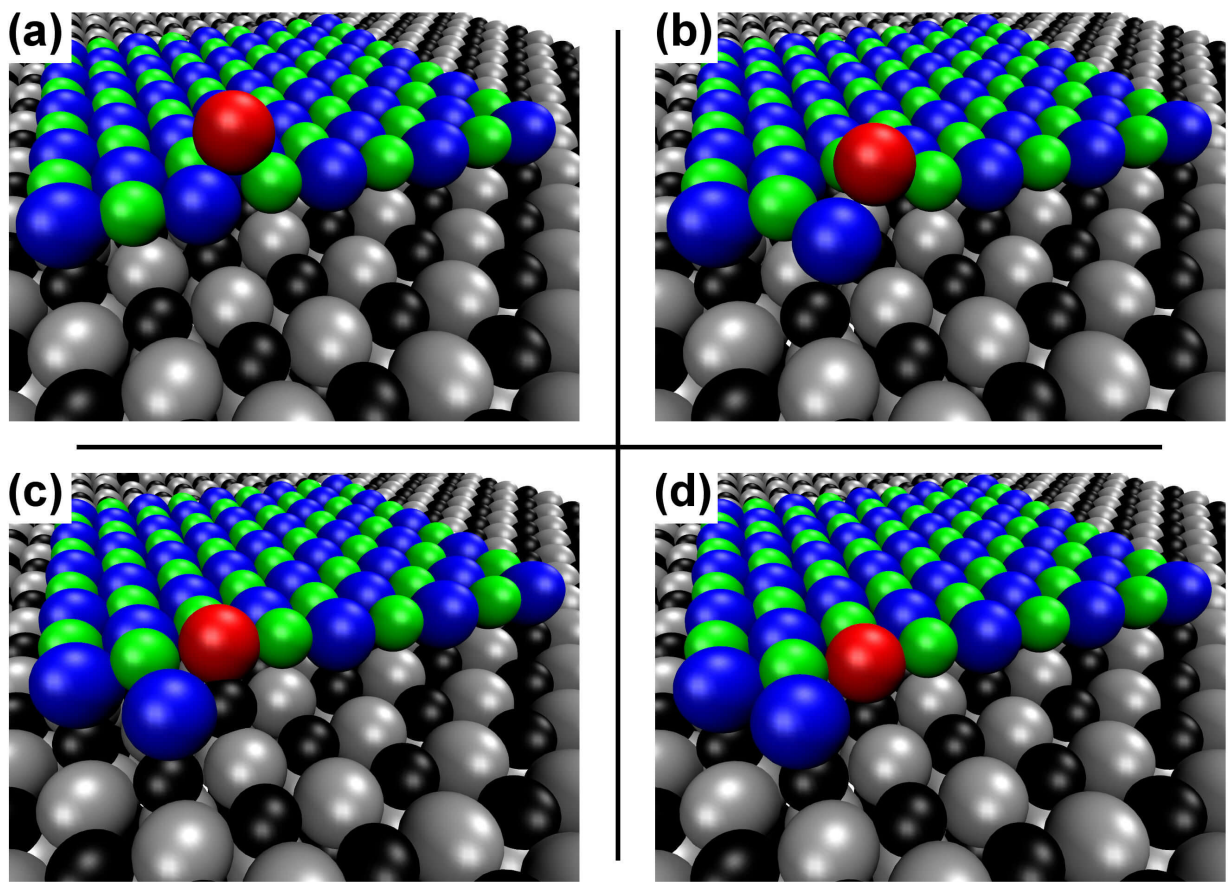

**Fig. 7**. The minimum-energy pathway for Ti$_{ad}$/Ti$_{isl}$ push-out/exchange on larger 9×9-atom TiN/TiN(001) islands based upon 0 K DFT+NEB calculations. ΔE in the upper panel is the difference in total energy with respect to the initial atomic configuration plotted as a function of the NEB plot is the in-plane distance (in arbitrary units) of the adatom from the island FFH edge site occupied by Ti$_{ad}$ prior to descent. The lower panels (a-d) illustrate the atomic configurations corresponding to the energies labeled in the upper panel.

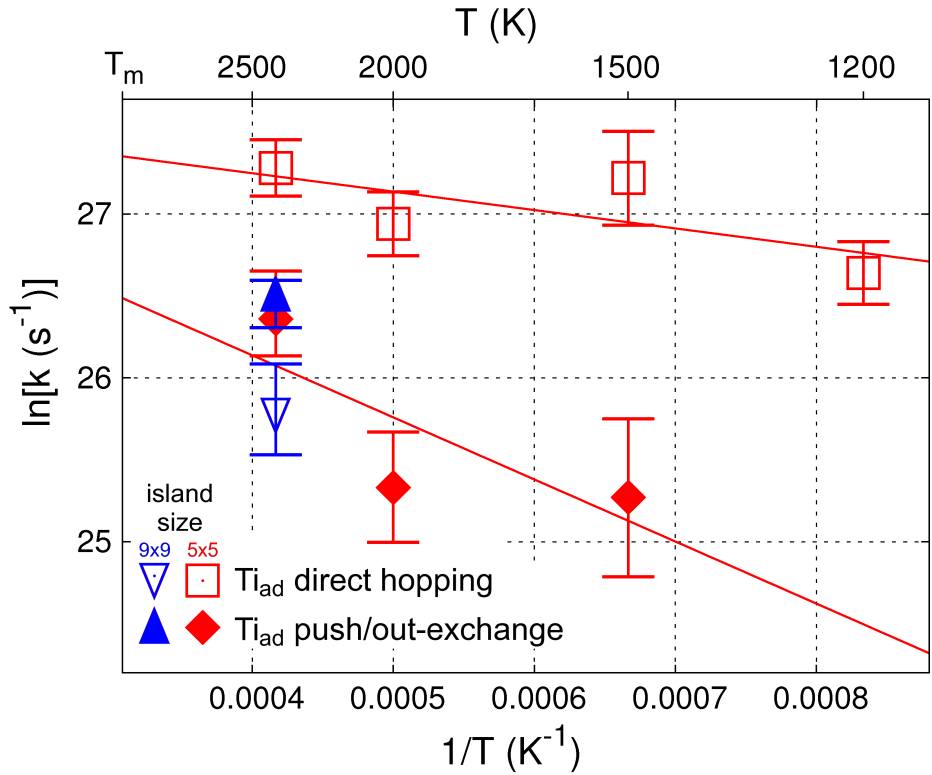

**Fig. 8**. Ti$_{ad}$ descent to the terrace rates ln[$k$] from FFH sites above 5×5- and 9×9-atom TiN(001) island step-edges plotted as a function of the inverse temperature 1/$T$. The melting point of TiN, $T_m$, is ~3250 K.

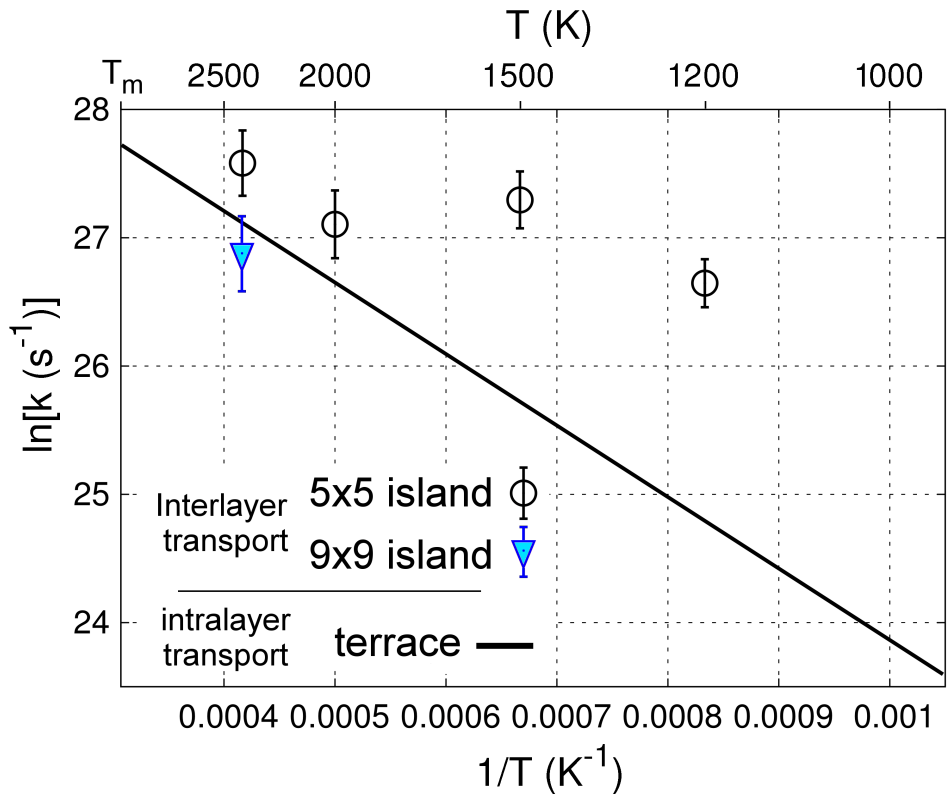

**Fig. 9**. Temperature dependence of total Ti$_{ad}$ descent to the terrace rates $k$ from FFH sites above 5×5- and 9×9-atom TiN(001) island step-edges vs. total Ti$_{ad}$ migration rates (solid line) on a TiN(001) terrace.

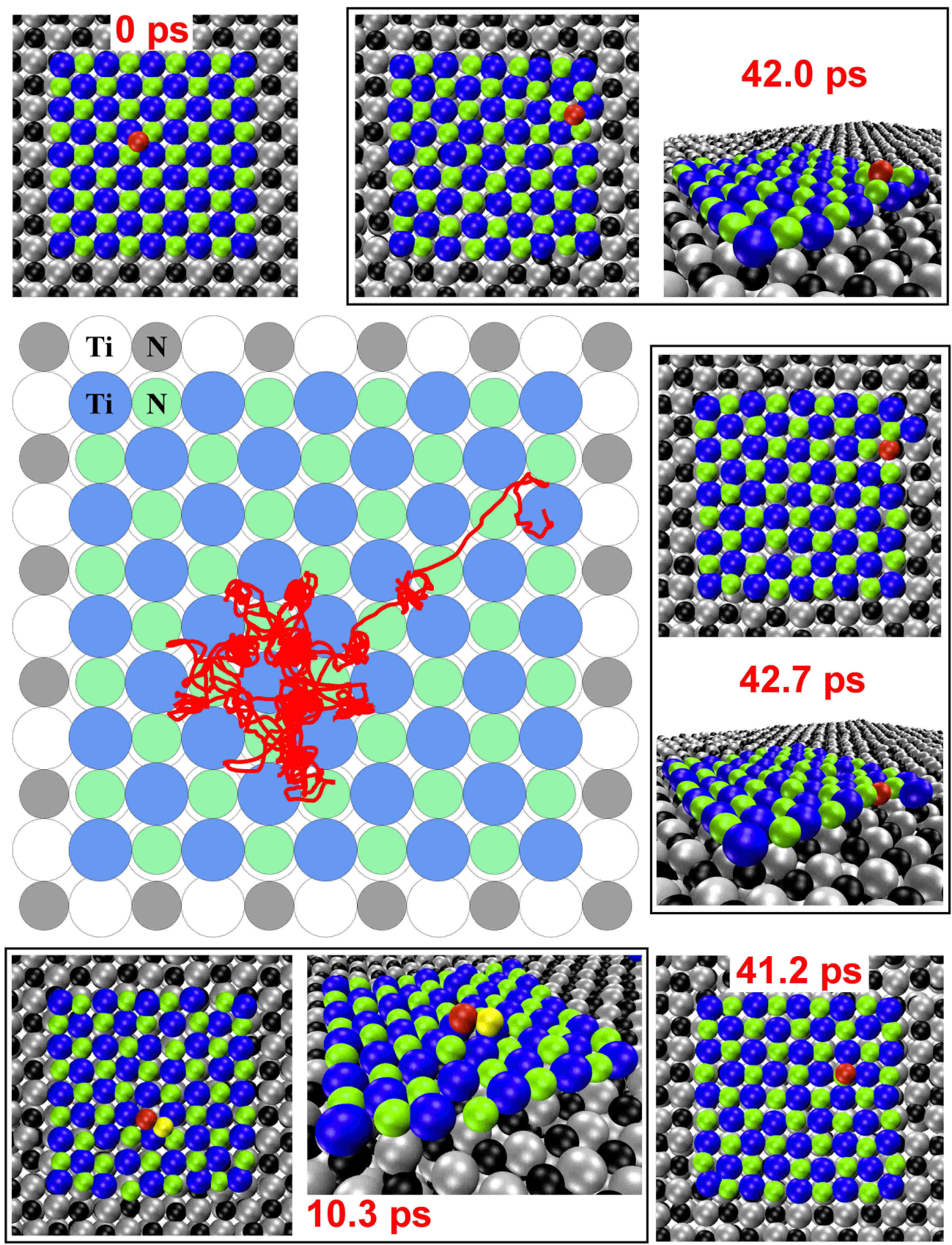

**Fig. 10**. Typical Ti adatom AIMD diffusion pathway (solid red line) on a 9×9-atom TiN/TiN(001) island at 2400 K. Snapshots illustrate events occurring during 42.7 ps simulation time. Red, blue, and silver spheres are the Ti adatom, Ti island atoms, and Ti terrace atoms, respectively. Yellow, green, and black spheres are a N island atom which is temporarily lifted by the Ti adatom onto the island, N island atoms, and N terrace atoms, respectively.

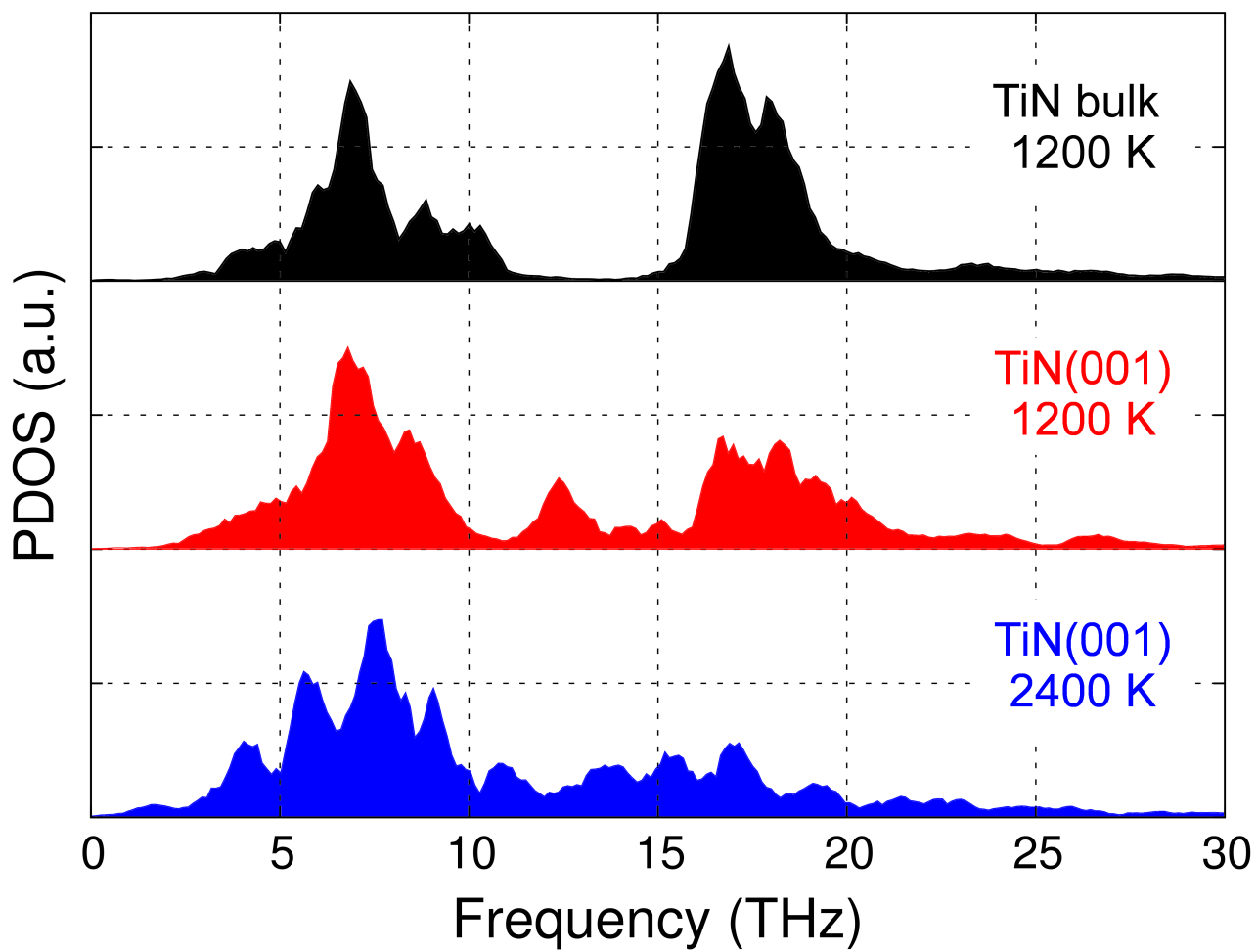

**Fig. 11**. TiN bulk and surface phonon densities of states at 1200 and 2400 K determined via AIMD velocity autocorrelation functions.

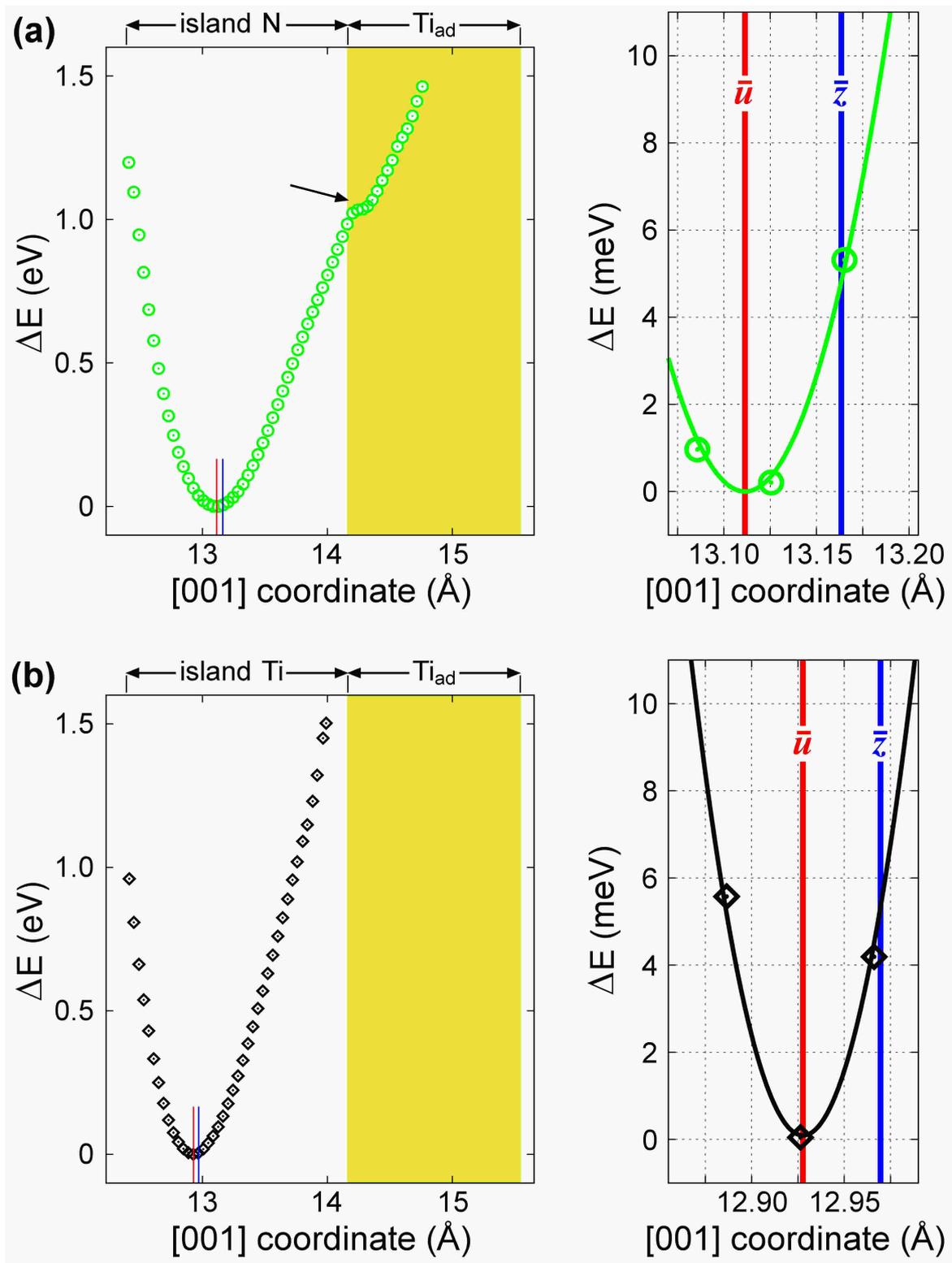

**Fig. 12**. Potential energy profiles along the surface normal direction of (a) N island and (b) Ti island atoms in 9×9 islands at 2400 K. Surface vibrations at 2400 K produce outward shifts (~0.05 Å) in island-atom [001] average coordinates $\bar{z}$ (vertical blue solid lines) with respect to their potential energy minima $\bar{u}$ (red solid lines) along the [001] direction. Panels on the right are magnifications of potential energy curves at the energy minima. Shaded yellow areas indicate the vertical coordinates of the Ti adatom recorded during 7 ps. For reference, N and Ti atoms in the surface TiN(001) layer (directly under the island) have average [001] coordinates of 10.85 Å.

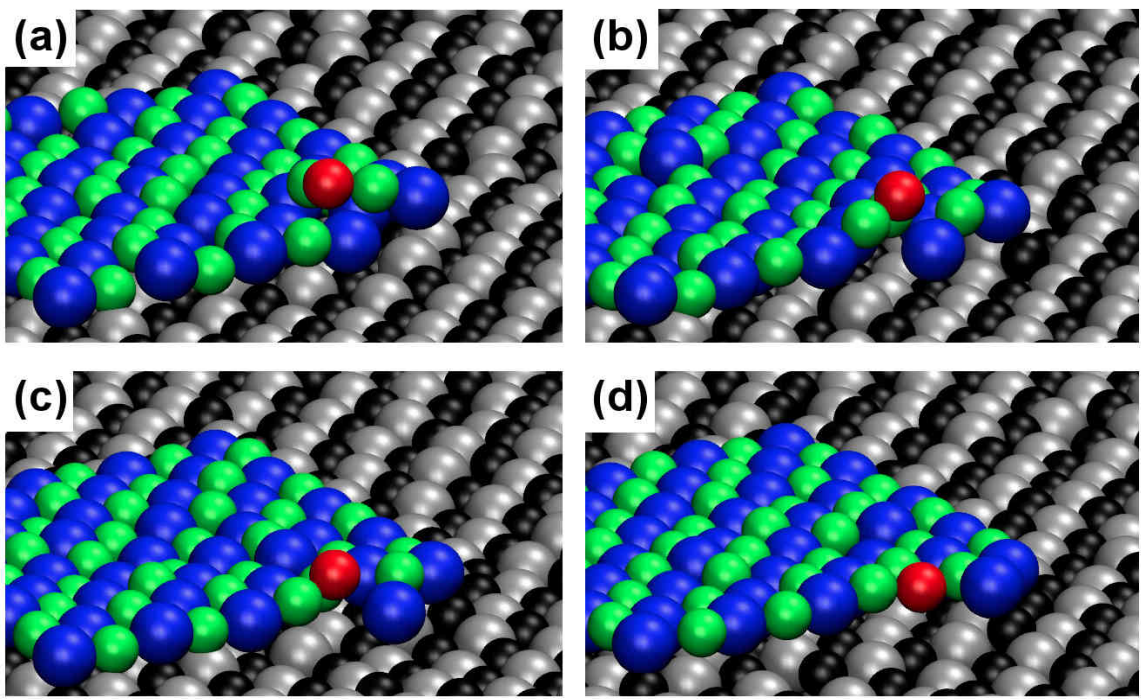

**Fig. 13**. Typical AIMD pathway for Ti adatom push-out/exchange descent from 9×9-atom <100>-faceted TiN/TiN(001) islands at 2400 K. Red, blue, and silver spheres are the Ti adatom, island atoms, and terrace atoms, respectively. Green and black spheres are N island and terrace atoms, respectively.